# Acoustic Sensing with Correlation and Coherent Detection using an Integrated Coherent Transceiver

André Sandmann
ADTRAN
Meiningen, Germany
asandmann@adva.com

Florian Azendorf
ADTRAN
Meiningen, Germany
fazendorf@adva.com

Saif Alnairat
ADTRAN
Meiningen, Germany
salnairat@adva.com

Michael Eiselt
ADTRAN
Meiningen, Germany
meiselt@adva.com

*Abstract*—Fiber optic sensing is becoming an important means to physically secure today's network infrastructure. However, a network-wide deployment of the monitors will require cost reduction of the interrogator system, which can only be achieved by highly integrated system components. In this contribution, we report on the use of an in-house designed single-chip coherent transceiver for acoustic fiber sensing. The transceiver on the basis of silicon photonics contains a high-speed dual-polarization IQ-modulator as well as a coherent receiver with balanced photodiodes and trans-impedance amplifiers, as defined by the OIF integrated coherent transmit-receive optical sub assembly (IC-TROSA) implementation agreement. The laser, used for transmission and as local oscillator, is provided external to the photonic integrated circuit and can be chosen according to the line-width requirements of the sensing system. The acoustic sensing demonstration is using a correlation-based optical time domain reflectometry with coherent detection. This method is able to detect, besides the amplitude information, the phase of the back-scattered signal, which has a significantly higher sensitivity to environmental effects on the fiber, like temperature and strain. As a proof of concept, sensing of an acoustic signal after a fiber span of 20 km is demonstrated by evaluating the obtained phase information, providing information on external dynamic events up to a frequency of 1.75 kHz.

## I. Introduction

Distributed acoustic sensing systems enable the detection of dynamic environmental effects that occur in the vicinity of a fiber. Such effects can be the vibration originating from e.g. an excavator that drives on soil above or close to a deployed fiber. This vibration translates into small changes of the refractive index and length of the fiber core, influencing the propagation delay. Measuring the time of flight of a probe signal that is sent into the sensor fiber and is reflected at e.g. a connector with a conventional OTDR is not sensitive enough to measure these small variations. By improving the spatial resolution and sensitivity with introducing correlation and pulse fitting, delay variations in the order of a few picoseconds can be measured [1]. Such delay variations can occur due to temperature changes of a fiber over a length in the order of kilometers. However, the sensitivity of this approach is not sufficient to measure variations that are applied on short fiber sections. In contrast, the phase information obtained from coherent detection is highly sensitive to environmental effects. Here, a temperature change of 1 Kelvin on a 1 centimeter unjacketed silica fiber section doped with germanium dioxide leads to a phase change of approximately 49 degrees. Furthermore, the application of 100 nanostrain on a 1-meter fiber section translates to a phase change of 53 degrees. Measuring acoustic signals can help to secure the fiber network infrastructure by e.g. preventing fiber cuts or detecting unauthorized physical access to this infrastructure. Furthermore, acoustic sensing can be used to monitor the environment of a fiber for a variety of applications [2,3].

However, phase-sensitive OTDR systems are quite costly, since they require an ultra-low phase noise laser to be able to measure phase changes from environmental effects rather than measuring the phase noise of the laser. Furthermore, such systems need to be capable of processing high data volumes in the initial signal processing steps, before applying data reduction strategies. In this contribution, the focus is on using an in-house designed highly integrated coherent transceiver based on silicon photonics to realize a coherent OTDR interrogator. The use of such a standardized telecom component and its high integration level offers a high energy efficiency, while the large volume of telecom components promises a low cost. In addition to coherent detection, correlation is used in order to improve the spatial resolution of the system. The interrogator concept is named coherent correlation OTDR (CC-OTDR) [3].

In this paper, we demonstrate the applicability of this integrated component for the detection of an acoustic signal after a fiber span of 20 km by analyzing the phase information of the reflected signals.

Section II discusses the structure of the in-house designed IC-TROSA. In Section III, the testbed and the acoustic sensing results are presented. The concluding remarks are provided in Section IV.

## II. Integrated Coherent Transceiver Structure

An integrated coherent transceiver (ICT) is a type of optical transceiver used in high-speed fiber optic communication systems. It combines the functions of a transmitter and a receiver into a single device. By utilizing coherent detection, the ICT can detect amplitude, phase and polarization of the optical signal, facilitating highly precise detection and decoding of the received data. The integration of the transmitter and receiver functions within the same device reduces the size, power consumption and cost of the overall system. ICTs are commonly used in long-haul and metro optical networks, where high-speed and high-capacity communication is required. They are also used in emerging applications such as data center interconnects and 5G wireless networks. In this work, the ICT is utilized in an acoustic sensing interrogator system. The in-house designed module used for this measurement is a digitally controlled board-mounted coherent transceiver with integrated drivers, transimpedance amplifiers, control and monitoring functions. The module is compliant with the OIF-IC-TROSA-01.0 Implementation Agreement [4], supporting C-band data rates up to 400 Gb/s, and symbol rates up to 64 GBaud. This module is based on surface mount ball grid array (BGA) configuration and can be used in various applications, including line cards and small form-factor pluggable transceivers.

The IC-TROSA's main interfaces include power supplies, a digital communications interface, specific input/output connections for alarm and control, high-speed RF contacts for I/Q data transmission and reception, and an optical fiber interface for laser connection [4]. In addition, the module offers an optional tunable amplified spontaneous emission (ASE) filter, which is integrated into the package. This option enhances overall integration and reduces circuit size. The tunable ASE filter can be controlled externally through analog interfaces that are also integrated into the BGA package. Control of the module is achieved through a standard two-wire interface. The module's transmitter control loops and performance monitoring are integrated into a single package, which provides a high level of integration of the transmitter-receiver drive chain.

The optical interface consists of edge couplers, which are equipped with V-grooves to facilitate passive mechanical alignment between the fibers and the silicon photonic chip of the module. These edge couplers include a spot size converter that is independent of polarization, and matches the spot size of a typical single-mode fiber.

A schematic of the IC-TROSA structure is shown in Fig. 2, compare box highlighted in yellow. The IC-TROSA provides a modulation function at the transmitter, employing a polarization multiplexed IQ modulator. Each IQ modulator is comprised of two inner nested Mach-Zehnder modulators with bias control, a phase shifter with phase control in the outer modulator, and an output power monitoring photodiode (MPD). The modulator function also includes linear RF driver amplifiers for each of the four Mach-Zehnder modulators. In the Mach-Zehnder arms, RF phase shifters are employed to rapidly alter the phase of the optical signals. The refractive index of the waveguide is changed through the use of carrier dispersion. This change affects both the real and imaginary components of the effective refractive index. Nevertheless, the waveguide's architecture is created in such a way that the imaginary component's change is much lower than that of the real component. Consequently, only a minimal amount of amplitude modulation is produced in addition to the intended phase modulation.

Optical power from the external laser source is divided into two parts with a polarization splitter rotator (PSR), and each part is independently modulated by a quadrature modulator. The resulting modulated signals in x and y polarization are then combined with a polarization beam combiner and output through an optical output fiber. The power in each of the two polarizations is independently monitored using MPDs and controlled via variable optical attenuators (VOAs). Similar to the RF phase shifters, waveguide diodes are used to implement the VOAs, which utilize the carrier dispersion effect to modify the real and imaginary parts of the waveguide's refractive index. However, the design is distinct because the imaginary part is more significantly affected than the real part. The in-phase and quadrature phase offset is controlled via an internal control algorithm executed and monitored by the IC-TROSA micro controller. Thermal phase shifters functioning as heating elements are located in close proximity to the optical waveguides, to utilize the thermal optical effect and modify the waveguide's refractive index. This modification solely affects the real component of the refractive index, resulting in a phase change without any alteration in the amplitude.

The received signal is split into two orthogonal polarizations using a PSR at the receiver side. The power in each of the two polarizations is independently monitored with MPDs. Moreover, VOAs are used at each path to attenuate and control the signal power. Each polarization is then delivered to a 90 degree hybrid mixer with differential optical outputs. The local oscillator input power is equally split and delivered to the two 90 degree hybrid mixers. To detect the signal, linear trans-impedance amplifiers are used in conjunction with photo-detectors consisting of 4 sets of differential detectors. Figure 1 shows the IC-TROSA chip on an evaluation board in the acoustic sensing testbed.

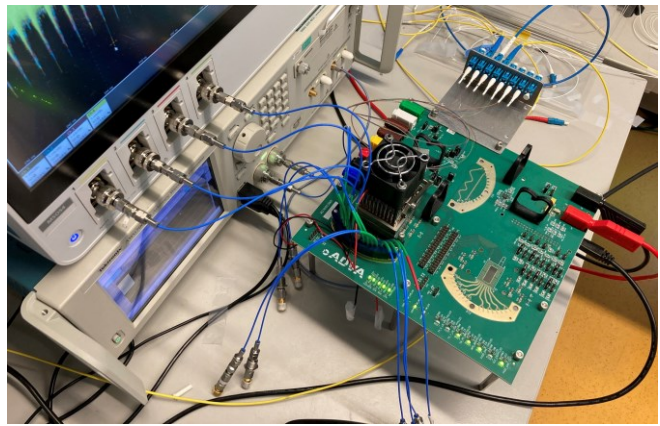

Fig. 1. IC-TROSA on evaluation board in the acoustic sensing setup

### III. TESTBED AND ACOUSTIC SENSING RESULTS

The IC-TROSA is integrated in the CC-OTDR testbed setup as it is depicted in Fig. 1 and schematically illustrated in Fig. 2 (compare box highlighted in yellow). Herein, continuous wave light of an ultra-low phase noise laser, exhibiting a Lorentzian line-width of less than 100 Hz, is fed to the transmitter input port (Tx in) of the IC-TROSA. One part of the laser power is used as the local oscillator (LO) for self-homodyne coherent detection. The other part is binary phase shift keying (BPSK) modulated at 125 MBaud in x-polarization with a pseudo-random binary sequence (PRBS) of length 8191, appended by a '-1' bit to obtain a balanced signal. This sequence is followed by a 27500 symbol long zero padding, which is designed to suppress the modulator output while the probe sequence propagates through the 20 km fiber. With this pattern with a repetition rate of 3.5 kHz, dynamic changes of the fiber up to a frequency of 1.75 kHz can be detected. The modulated pulse sequence is amplified in an erbium doped fiber amplifier (EDFA), and the corresponding amplified spontaneous emission noise is limited by an optical bandpass filter (OBF). Finally, the amplified pulse sequence is sent out to the sensor fiber using a circulator. Signal parts that are back-scattered and reflected from this fiber are transferred to the receiver input (Rx), where they are mixed with the LO for each polarization separately. Subsequently, the in-phase and quadrature components are extracted, and the four orthogonal electrical field components are recorded with an oscilloscope. Further signal processing is performed offline. Here, the received signals are cross-correlated with the transmitted pulse sequence to achieve a pulse compression and SNR improvement.

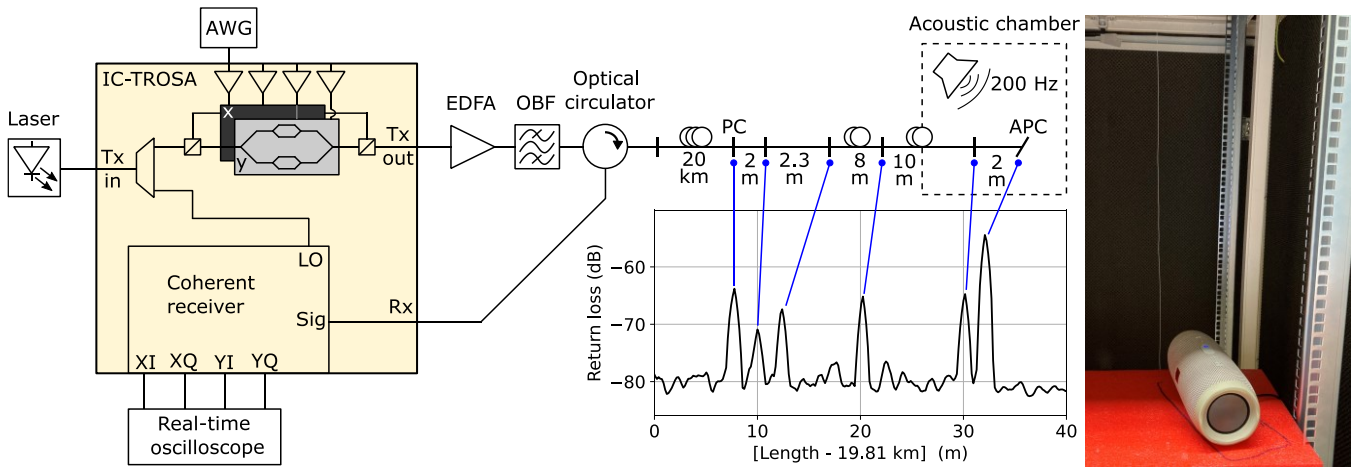

Fig. 2. Acoustic sensing testbed schematic using the IC-TROSA

The sensor fiber consists of a 20 km launch fiber followed by a series of patch cords with different lengths, being linked with physical contact (PC) connectors. An open angled physical contact (APC) connector with a white dust cap is used to terminate the sensor fiber, providing a return loss of 55 dB. In the return loss trace in Fig. 2, the reflections from the connectors are shown as peaks, indicating a 10 to 25 dB higher reflectivity compared to the Rayleigh back-scattering level. Figure 3 shows the return loss trace of the complete fiber. The peaks at 0 km and 20 km are reflections from fiber connectors, where the peak at the fiber end comprises all reflections visible in the return loss trace in Fig. 2. As in conventional OTDR systems, the Rayleigh back-scattering leads to an interference pattern originating from the coherent superposition of back-scattered light from different locations. The resulting fading pattern is characteristic for a specific fiber being probed at one wavelength and at constant environmental conditions, since inhomogeneities due to the amorph structure of the doped silica fiber core lead to concentrated scattering regions of the fiber. Thus, this pattern can be viewed as the fiber fingerprint. Furthermore, the fingerprint envelope indicates a round-trip attenuation of 0.38 dB/km (or 0.19 dB/km single-pass attenuation), as shown by the blue line in Fig. 3. The noise floor after the last reflection originates from correlation sidelobe noise, where the connector reflections are the main contributor.

A fiber fingerprint example, showing individual transmission frames of a 40-meter fiber section in a Rayleigh back-scattering region at 1.1 km into the sensor fiber, is shown in Fig. 4. It highlights that the maxima locations (resulting from constructive interference) and minima locations (originating from destructive interference) are stable over time in steady environmental conditions. It is worth noting, that the low signal-to-noise ratio at minima locations prevents a phase data analysis. Thus, fading mitigation strategies by e.g. transmitting at multiple frequencies have been proposed [5].

In the testbed setup, the last 2 meters of the 10-meter fiber section as well as the subsequent final 2-meter long PC to APC patch cord are located in an acoustically isolated chamber. Herein, a speaker is used to apply a 200 Hz sinusoidal acoustic signal to the sensor fiber, which is hanging from the ceiling of that chamber, see picture in Fig. 2. By comparing the received phase information from the input connector of the 10-meter fiber section to the phase of the open APC connector, the resulting phase difference signal reflect the changes from environmental effects on this fiber section. The measured phase difference signal is depicted in Fig. 5, clearly showing variations with a peak-peak variation of approximately 0.9 rad at the applied 200-Hz acoustic signal (black line). With no acoustic signal present, the phase difference shows significantly lower variations (blue line). The remaining variations are a combination of noise sources from the interrogator, like laser phase noise and thermal noise, and noise from environmental effects, like remaining acoustic noise, airflow and thermal changes. These experimental results highlight that such an integrated transceiver is well suited for acoustic sensing.

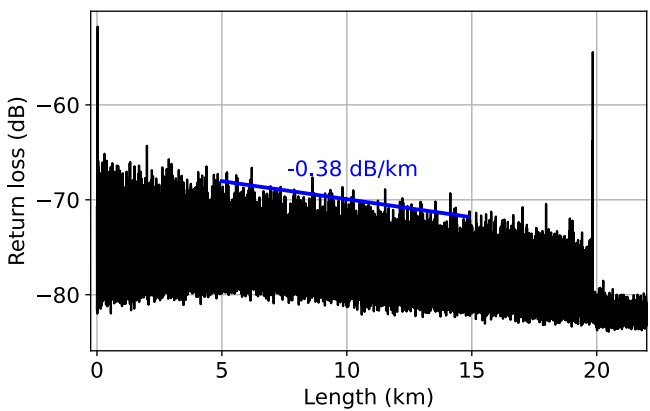

Fig. 3. Fingerprint of the 20 km sensor fiber

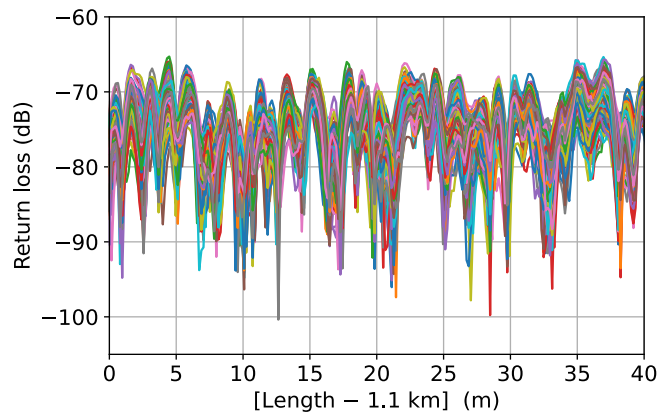

Fig. 4. Measured fingerprints with each color showing an individual frame of a 40-meter fiber section in the Rayleigh back-scattering region at 1.1 km

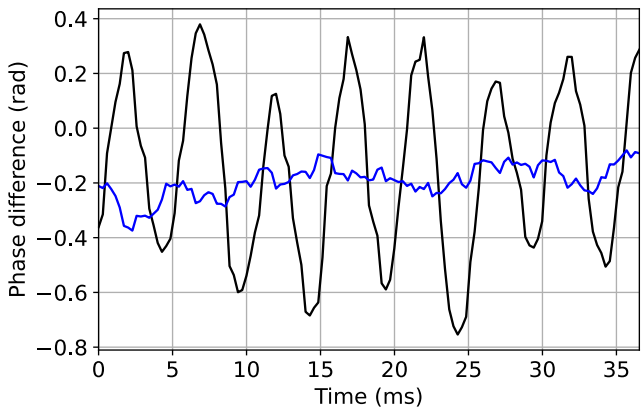

Fig. 5. Measured phase difference signal with (black line) and without (blue line) applying a 200-Hz acoustic signal

## IV. Conclusion

In this paper, a single-chip highly integrated coherent transceiver, designed for telecom applications, has been used in a coherent correlation OTDR interrogator for acoustic sensing. As a proof of concept, the detection of an acoustic signal applied to a fiber section via a speaker after a fiber span of 20 km has been demonstrated based on evaluating the phase data. The results highlight that such an highly integrated and energy efficient component is suitable for acoustic sensing applications.

## Acknowledgment

This work has received funding from the Horizon Europe Framework Programme under grant agreement No 101093015 (SoFiN Project).